# Beyond Tools: Generative AI as Epistemic Infrastructure in Education

Bodong Chen, University of Pennsylvania

**Abstract:** AI systems are increasingly embedded in practices where humans have traditionally exercised epistemic agency, the capacity to actively engage in knowledge formation and validation. This paper argues that understanding AI's impact on epistemic agency requires analyzing these systems as *epistemic infrastructures* rather than as neutral tools. Drawing on theories of technological mediation and distributed cognition, I advance a framework that foregrounds how AI systems reconfigure the conditions under which epistemic agency can be exercised. The framework specifies three analytical conditions: affordances for skilled epistemic actions, support for epistemic sensitivity, and implications for habit formation. I apply this framework to AI systems deployed in education, a domain where epistemic agency is both professionally essential and ethically significant. Analysis of AI lesson planning and feedback tools reveals patterns of *epistemic substitution*: while useful for efficiently handling teaching tasks, these systems perform cognitive operations without sustaining skilled epistemic actions, epistemic sensitivity, or virtuous habit formation, potentially preventing the cultivation of professional judgment that relies on these practices. The findings contribute to philosophical debates about AI and human agency by specifying mechanisms through which infrastructural embedding shapes epistemic possibilities, and offer design principles for AI systems that sustain rather than supplant human epistemic agency.

## 1. Introduction

Artificial intelligence (AI) systems are increasingly performing epistemic work traditionally done by humans—generating knowledge claims, evaluating evidence, synthesizing information, and rendering judgments.[1] This shift represents more than technological change; it constitutes a transformation in how knowledge is produced, validated, and maintained within social institutions. Yet analysis of how AI systems reshape human epistemic agency remains underdeveloped. Most analyses focus on AI's outputs (their accuracy, bias, or reliability) rather than on how AI integration restructures the epistemic practices through which humans engage with knowledge. This paper argues that examining AI as *epistemic infrastructure* rather than as tool reveals distinctive challenges for human knowledge practices that instrumental framings obscure.

*Epistemic agency*—the capacity to actively engage in and control one's own belief formation and validation processes (Ahlstrom-Vij, 2013)—depends not only on individual cognitive capabilities but on the infrastructural conditions that enable or constrain epistemic actions. Philosophy of technology has long recognized that technologies are not neutral instruments but actively mediate human experience and action (Ihde, 1990; Verbeek, 2008). Technologies shape what becomes visible, what questions appear reasonable, and what forms of engagement seem possible (Feenberg, 2012). When AI systems become embedded in epistemic practices, they do not simply assist human agents but reconfigure the conditions under which epistemic agency can be exercised.

---

[1] AI and generative AI are not distinguished in this paper. Given the centrality of generative AI in current discourse, AI refers primarily to generative AI in this paper unless explicitly stated otherwise.

This reconfiguration is consequential in domains where epistemic agency is both professionally essential and ethically significant, such as medicine (Herzog & Branford, 2025) and education (Markauskaite & Goodyear, 2017). In education, teaching is fundamentally epistemic work: educators evaluate evidence of student understanding, make judgments about pedagogical approaches, and develop situated knowledge through sustained engagement with learners and curricula (Shulman, 1987). When AI systems mediate these practices—generating lesson plans, providing feedback on student work, suggesting instructional strategies—they do not merely augment teacher capabilities. Rather, they restructure the epistemic labor through which teaching expertise develops and is exercised. The question is not simply whether AI produces accurate outputs, but whether AI-mediated practice supports or undermines the epistemic agency that constitutes professional teaching.

In this paper, I advance a framework for analyzing AI systems as epistemic infrastructures, drawing on theories of technological mediation (Ihde, 1990; Verbeek, 2008), distributed and extended cognition (Giere, 2002; Heersmink & Knight, 2018; Hutchins, 1991), and scholarship on epistemic agency and infrastructure (Coeckelbergh, 2022; Marin, 2022). The framework foregrounds three analytical dimensions: (1) affordances for skilled epistemic actions, (2) support for epistemic sensitivity, and (3) implications for habit formation. Applied to AI systems in education, these dimensions reveal how technologies marketed as supporting teachers may instead perform the cognitive operations through which professional knowledge develops, a pattern called *epistemic substitution*. Unlike cognitive offloading generally, epistemic substitution is problematic specifically because the operations being delegated are those whose exercise constitutes expertise development and establish authority (Mehdizadeh & Hilbert, 2025). The analysis contributes to broader philosophical debates about AI's impact on human agency by specifying mechanisms through which infrastructural embedding, not merely tool use, shapes epistemic possibilities.

The paper proceeds as follows. I first examine how theories of technological mediation and distributed cognition illuminate AI's distinctive character as epistemic infrastructure. I then elaborate the analytical framework, specifying conditions that infrastructures must support to foster epistemic agency. The framework is applied to two AI systems widely deployed in educational settings—an AI lesson planner and an AI feedback tool—examining how these systems mediate teaching practices where educators have traditionally exercised considerable epistemic agency. The analysis reveals that current AI systems inadequately support skilled epistemic actions, insufficiently foster epistemic sensitivity, and risk cultivating problematic habits of epistemic substitution. I conclude by discussing theoretical implications and offering design principles for AI systems that sustain human epistemic agency.

## 2. Theoretical Framework

### 2.1 Technological Mediation and Human-Technology Relations

Central to understanding how AI systems reshape epistemic practices is the concept of *technological mediation*: Technologies do not function as neutral instruments but actively shape how humans perceive the world and act within it (Verbeek, 2008). This insight challenges instrumentalist views that treat technologies as mere means to human ends, revealing instead how technologies become constitutive elements of human experience.

Ihde's (1990) phenomenological analysis distinguishes several forms of human-technology relations. In *embodiment relations*, technologies become transparent extensions of human capabilities, such as eyeglasses through which one sees. The technology recedes from awareness as it amplifies bodily capacity. In *hermeneutic relations*, technologies provide representations requiring interpretation: thermometers displaying temperature, maps depicting terrain. Here the technology remains present as something to be read. In *alterity relations*, humans engage technologies as quasi-others possessing a degree of independence, such as ATMs that respond or robots that navigate. Finally, in *background relations*, technologies shape conditions for action while occupying a background position in human experience, as when climate control systems maintain temperature, electrical grids supply power, or algorithms curate content feeds.

Each relational form configures human perception and action differently, shaping what becomes visible, possible, and meaningful. Insulated gloves in embodiment relation extend one's capacity to handle cold weather; a thermometer in hermeneutic relation presents temperature as a reading to be interpreted; a smart thermostat in alterity relation learns preferences and adjusts with apparent autonomy; a central climate control system in background relation maintains the thermal environment for the whole building. The philosophical significance lies not in the technology itself but in how it mediates the human-world relation.

AI systems present distinctive challenges to these established categories. Unlike mechanical tools or conventional software, AI technologies perform interpretive and inferential operations previously exclusive to human cognition. When an AI system generates a lesson plan or provides feedback on student work, it does not simply extend human capability (embodiment) nor merely present information for interpretation (hermeneutic). The AI selects, synthesizes, evaluates, and generates, performing cognitive operations that constitute knowledge work. Verbeek (2008) identifies this as *composite intentionality*, when intentionality of both human beings and technological artifacts gets combined, blurring traditional boundaries between user and tool. The human using an AI system does not simply employ a tool to accomplish a pre-existing goal; the AI's design, processing, and outputs actively shape what goals appear reasonable, what approaches seem viable, and what counts as adequate performance.

This technological mediation carries profound implications for epistemic practices. Technological design embeds specific values, interests, and configurations of power (Feenberg, 2012). Technologies privilege certain actions while foreclosing others, thereby determining the "range of choices" available to users. Design decisions—about interfaces, defaults, what information to surface or suppress—shape practice in ways that often escape critical scrutiny precisely because they become infrastructural, taken-for-granted backgrounds to action. When AI systems mediate epistemic practices, they do not merely make these practices more efficient; they reconfigure what these practices entail, who performs them, and what forms of knowledge become possible or marginalized.

## 2.2 Epistemic Agency and Its Distribution

*Epistemic agency* refers to the capacity to actively engage in and control one's own belief formation and reflection processes, rather than passively accepting information (Ahlstrom-Vij, 2013). This capacity manifests through skilled actions, such as gathering evidence, evaluating sources, weighing competing interpretations, forming justified beliefs, and revising beliefs in light of new information. Epistemic agency is not merely cognitive but practical, exercised through what we do when pursuing knowledge, shaped by the epistemic virtues we cultivate and the norms of inquiry we follow (Elgin, 2013; Setiya, 2013).

Crucially, epistemic agency is not confined to individual minds. Theories of distributed cognition demonstrate that cognitive processes routinely extend across persons, artifacts, and environmental structures (Hutchins, 1991). Hutchins's analysis of ship navigation reveals cognition distributed across crew members, instruments, charts, and established procedures, forming a cognitive system exceeding any individual's knowledge. Similarly, Knorr Cetina (1999) and Giere (2002) show how scientific knowledge production occurs through "epistemic cultures" where cognition distributes across researchers, instruments, databases, and collaborative practices. Knowledge often resides not in individual minds but in socio-material assemblages.

These insights illuminate how epistemic agency operates within *epistemic infrastructures*, defined as the frameworks and systems through which knowledge is created, circulates, and can be validated (Coeckelbergh, 2022; Marin, 2022). Such infrastructures encompass the cultural and structural supports that enable epistemic operations, from libraries and universities to search engines and peer review systems, and now to AI systems granted authority to arbitrate facts (Mehdizadeh & Hilbert, 2025). They shape what knowledge is accessible, what methods of inquiry are available, and what standards of justification apply.

The relationship between individual epistemic agency and infrastructural conditions is complex. Infrastructures can enhance epistemic agency by providing resources, tools, and collaborative structures that extend individual capabilities. A well-designed research library supports scholarship; scientific instrumentation enables observations otherwise impossible; collaborative platforms facilitate knowledge exchange. However, infrastructures can also constrain or undermine epistemic agency. In Heersmink and Knight's (2018) analysis of "extended cognitive systems," integration between human and artifact varies along dimensions including accessibility, trust, and transparency. Dense integration constitutes genuine cognitive extension; shallow integration merely scaffolds internal processes. They note that reliance on certain technologies may result in atrophied capacities when those technologies become unavailable.

Consider weather forecasting as epistemic infrastructure. Edwards (2010) demonstrates that climate knowledge depends on vast sociotechnical networks where data and models are mutually constitutive since raw observations require models for interpretation, calibration, and integration into coherent datasets. Climate scientists, meteorologists, and ordinary citizens interact with these networks of satellites, sensors, computational models, and forecasting services. Different actors relate to this infrastructure differently, but the system generally supports formation of context-responsive beliefs that guide appropriate action. Crucially, the infrastructure enables rather than replaces human judgment; the models produce forecasts, but humans interpret their significance for particular contexts and decide how to act. Social media platforms, by contrast, often function as problematic epistemic infrastructures. Optimized for engagement rather than justified belief formation, these platforms prioritize emotionally provocative content while algorithms personalize information environments in opaque ways (Marin, 2022). The delegation of epistemic operations to such systems can hijack epistemic agency, where systems exploit rather than support human knowledge practices (Dorsch, 2022).

## 2.3 AI as Distinctive Epistemic Infrastructure

AI systems are distinguished from previous epistemic infrastructures not primarily by scale or speed but by the kind of cognitive work they perform. Where previous technologies mediated access to information or structured its presentation, AI systems perform operations that were previously internal to human epistemic processes, such as selecting what information matters, evaluating quality, synthesizing across sources, generating novel content. This represents a qualitative shift in technological mediation.

Educational technologies illustrate the contrast. Historically, textbooks mediate knowledge processes but remain transparent to critical examination: readers can evaluate arguments, check sources, identify assumptions. Learning management systems structure interaction but leave interpretive work to humans: teachers still read student submissions, form judgments, provide feedback. Even search engines, while shaping what information surfaces, preserve the human role in navigating, evaluating, and synthesizing results. AI systems, however, can perform these evaluative and synthetic operations themselves. An AI feedback system does not merely present student writing for teacher review; it evaluates the writing and generates feedback, sometimes without keeping the teacher in the loop. An AI lesson planner does not simply provide resources for teachers to arrange; it produces complete instructional sequences and materials, sometimes with pedagogical orientations unchecked (Chen et al., 2025).

This shift creates conditions for *epistemic substitution*, where AI performs the cognitive operations that both reflect and develop professional expertise. The concept differs from cognitive offloading generally, as we routinely delegate cognitive tasks to technologies without concern. What makes epistemic substitution problematic is that the delegated operations are precisely those through which professional knowledge develops. Teachers develop expertise in assessing student understanding through sustained practice reading and responding to student work; pattern recognition, contextual sensitivity, and pedagogical judgment emerge through this engagement. When AI systems substitute for rather than support this cognitive work, they may prevent the formation of the very expertise they appear to enhance. The resulting cognitive arrangement is brittle: practitioners become dependent on systems that, when unavailable, reveal atrophied rather than augmented capabilities.

From a virtue epistemology perspective, this poses a fundamental challenge (Heersmink & Knight, 2018). Educational technologies should ideally cultivate epistemic virtues: intellectual curiosity, careful evaluation, epistemic humility, responsibility for one's beliefs. Yet when AI systems perform the cognitive work through which these virtues develop, they may inadvertently undermine epistemic formation. A student who relies on AI to evaluate sources never develops source evaluation skills. A teacher who relies on AI-generated feedback may never acquire the tacit knowledge that emerges from direct engagement with student work.

## 2.4 Education as Paradigmatic Domain

Education offers a paradigmatic domain for examining AI's impact on epistemic agency for several reasons. First, education is fundamentally epistemic work. Unlike domains where AI might enhance efficiency without affecting core practices, education's central activities—teaching, learning, assessment—are constituted by epistemic operations. Teachers form beliefs about student understanding, evaluate evidence of learning, make judgments about pedagogical effectiveness.

Students develop knowledge, learn to evaluate claims, cultivate epistemic virtues. AI integration thus affects not peripheral functions but the practices that define educational activity.

Second, epistemic agency in education carries distinctive ethical weight. Developing students' capacity for independent thought, critical evaluation, and intellectual autonomy has long been recognized as central to education's purpose, particularly in democratic societies (Barzilai & Chinn, 2018; Dewey, 1916). Teachers exercise professional judgment that shapes students' intellectual development. When AI systems reconfigure these epistemic practices, the stakes extend beyond efficiency or convenience to questions about what kind of epistemic agents education seeks to nurture.

Third, AI is being rapidly integrated into education systems in ways that exemplify broader patterns. The educational technology market now includes AI-powered tools for lesson planning, assessment, content generation, tutoring, and administrative functions (Celik et al., 2022; Kasneci et al., 2023). These systems are often marketed as reducing teacher workload or personalizing student learning. Yet examining how these systems actually mediate practice is critical for education.

Teachers occupy a distinctive position within education's epistemic infrastructures. While teachers have historically faced constraints on professional autonomy through standardized curricula, accountability measures, and administrative control (Ingersoll, 2003), they have generally maintained considerable epistemic agency in the micro-practices of teaching: making instructional decisions, reading and interpreting student work, developing practical knowledge through direct engagement (Shulman, 1987). These practices are now precisely what AI systems aim to mediate if not fully substitute. The question is not whether AI can perform these operations well enough, but whether AI-mediated practice supports or undermines the epistemic agency through which teaching expertise develops and is exercised.  Conducting such analysis will potentially reveal patterns that may apply across professional domains where situated, relational knowledge matters.

## 3. Analytical Framework

The literature analysis established that AI systems function as epistemic infrastructures that mediate human knowledge practices in consequential ways. To examine how specific AI systems shape epistemic agency, this paper conducts what information systems scholars term a technical investigation (Friedman & Hendry, 2019), defined as a systematic examination of how technological systems embody particular values through their design features, interface structures, and interaction patterns. Rather than studying user behavior, this approach analyzes the technologies themselves—their affordances, constraints, defaults, and workflows—to reveal how they configure possibilities for human action.

This analytical strategy has philosophical precedent. Just as Ihde's (1990) postphenomenological analyses examine how specific technologies mediate human-world relations, and Feenberg's (2012) critical investigation of how design choices embed values and power relations, technical investigation treats technologies as philosophical data. The question is not primarily how users actually behave (an empirical matter) but what forms of engagement the technology makes possible, encourages, or forecloses (a structural matter). Interface designs, default settings, and interaction flows encode assumptions about what users should do and what counts as successful use. Examining these features reveals the epistemic possibilities that AI systems create for their users.

## 3.1 Three Conditions for Epistemic Agency

To structure this investigation, I employ an analytical framework developed by Marin (2022) for examining digital environments as epistemic infrastructures. Marin identifies three conditions that environments must support if they are to foster rather than undermine epistemic agency: affordances for *skilled epistemic actions*, support for *epistemic sensitivity*, and implications for *habit formation*. While Marin developed this framework to analyze social media platforms, these conditions apply productively to AI systems in professional contexts. Each condition illuminates a different mechanism through which technological design shapes users' epistemic possibilities.

*Skilled epistemic actions* refers to the range and quality of epistemic operations that an infrastructure enables users to perform. Epistemic agency manifests through action: gathering evidence, evaluating sources, weighing interpretations, forming and revising judgments. An infrastructure that supports epistemic agency provides users opportunities to exercise these capacities in meaningful ways, drawing on their existing expertise and developing new competencies. The analytical question is: What epistemic actions does the AI system afford? Does it enable users to engage in complex, skilled operations, such as questioning, refining, critiquing, contextualizing, or does it reduce engagement to simplified choices (accept/reject, use/discard)? When systems limit user control to binary responses to AI-generated outputs, they constrain the epistemic work users can perform.

*Epistemic sensitivity* refers to users' capacity to recognize situations that warrant epistemic attention, moments when critical evaluation, verification, or deeper engagement is appropriate. Epistemic agency requires not only the ability to perform skilled actions but the disposition to recognize when such actions are called for. Infrastructures can cultivate or erode this sensitivity. Systems that present outputs as authoritative, that lack transparency about uncertainty or limitations, or that optimize for seamless acceptance may condition users to engage uncritically. The analytical question is: Does the AI system's design prompt users to exercise judgment, or does it encourage passive acceptance? Do interface features signal when scrutiny is warranted, or do they present AI outputs as settled conclusions?

*Habit formation* refers to how repeated interactions with an infrastructure shape users' long-term epistemic dispositions. Through sustained use, interaction patterns become automatic, forming habits that either reinforce or erode epistemic agency. Systems designed to streamline decision-making may foster habits of dependence; systems that build in moments for reflection may cultivate habits of critical engagement. The analytical question is: What epistemic habits does sustained use of the AI system likely produce? Does the system's workflow encourage deliberation and independent judgment, or does it reinforce patterns of deference to AI-generated outputs?

These three conditions are interconnected. Limited affordances for skilled action make it difficult to exercise epistemic sensitivity even when users recognize its importance. Eroded sensitivity reduces occasions for skilled action even when affordances exist. And habits formed through constrained, unreflective interaction become self-reinforcing, further diminishing both skilled action and sensitivity over time. Analyzing AI systems through these conditions reveals how design choices accumulate into epistemic consequences.

## 3.2 Cases and Data Sources

I apply this framework to two AI systems widely adopted in educational settings: MagicSchool AI, a platform with a comprehensive suite of AI tools for tasks like lesson planning, and Brisk, an AI-powered writing feedback system. These cases were selected for three reasons. First, both systems are commercially successful and widely adopted, representing mainstream rather than experimental AI integration in education. Second, they mediate essential practices, i.e., lesson planning and assessment, where teachers have traditionally exercised substantial epistemic agency through direct engagement with pedagogical materials and student work. Third, sufficient public information exists to conduct technical investigation: promotional materials, platform documentation, publicly accessible features, and user-generated demonstration videos.

The analysis draws on multiple data sources: system interfaces and interaction flows observed through platform access and public demonstrations; promotional materials articulating intended use cases and value propositions; platform documentation describing features and workflows; and published analyses of these systems. By triangulating across these sources, the investigation examines not only what these systems technically enable but how they position users and frame the work of teaching.

The analysis that follows examines each system through these three conditions, revealing how AI tools marketed as supporting teacher practice may instead inadvertently instantiate epistemic substitution, performing operations whose exercise is constitutive of teaching expertise.

## 4. AI Systems as Epistemic Infrastructure in Education

### 4.1 Case 1: MagicSchool AI

MagicSchool AI exemplifies the efficiency-focused design prevalent in educational AI. The platform generates complete lesson plans from minimal input—grade level, subject, topic, and duration—positioning speed as its primary value proposition. Marketing materials emphasize that the tool "simplifies the process" of lesson planning, promising to help teachers "save time" that can be "redirected toward instruction." The platform offers various generators for different lesson formats, each producing structured plans that include learning objectives, activities, assessments, and materials lists. Figure 1 illustrates the interface.

*Skilled epistemic actions.* Lesson planning traditionally involves complex epistemic work, including interpreting curriculum standards, selecting and sequencing content, anticipating student difficulties, designing activities that scaffold understanding, and aligning assessments with learning goals (Shulman, 1987). This work draws on teachers' content knowledge, pedagogical expertise, and contextual understanding of their students. MagicSchool restructures this epistemic labor. The platform does offer customization; teachers can specify parameters, edit generated content, and iterate on outputs. However, the interface design rewards efficiency over refinement. The workflow encourages users to generate plans quickly and move on, with no structural incentives for the deeper pedagogical reasoning that characterizes expert lesson design. Teachers can engage in skilled epistemic action, but doing so requires working against the platform's primary affordances rather than with them. The system thus allows but does not cultivate sophisticated epistemic engagement.

*Epistemic sensitivity.* The platform's generation process remains opaque: teachers cannot inspect how the AI selects content, sequences activities, or aligns with standards. This opacity creates conditions for what research on automated decision-making systems identifies as blind trust and over-reliance on automated outputs (Klingbeil et al., 2024). MagicSchool presents generated plans as valid educational content without foregrounding questions about pedagogical soundness, curricular coherence, or contextual appropriateness. The interface includes no prompts for critical evaluation, no indicators of uncertainty, no mechanisms for teachers to assess whether outputs meet epistemic standards for lesson quality. Teachers who wish to maintain epistemic vigilance must supply their own evaluative frameworks and invest additional effort that the platform's efficiency orientation discourages. The design structurally rewards acceptance over scrutiny.

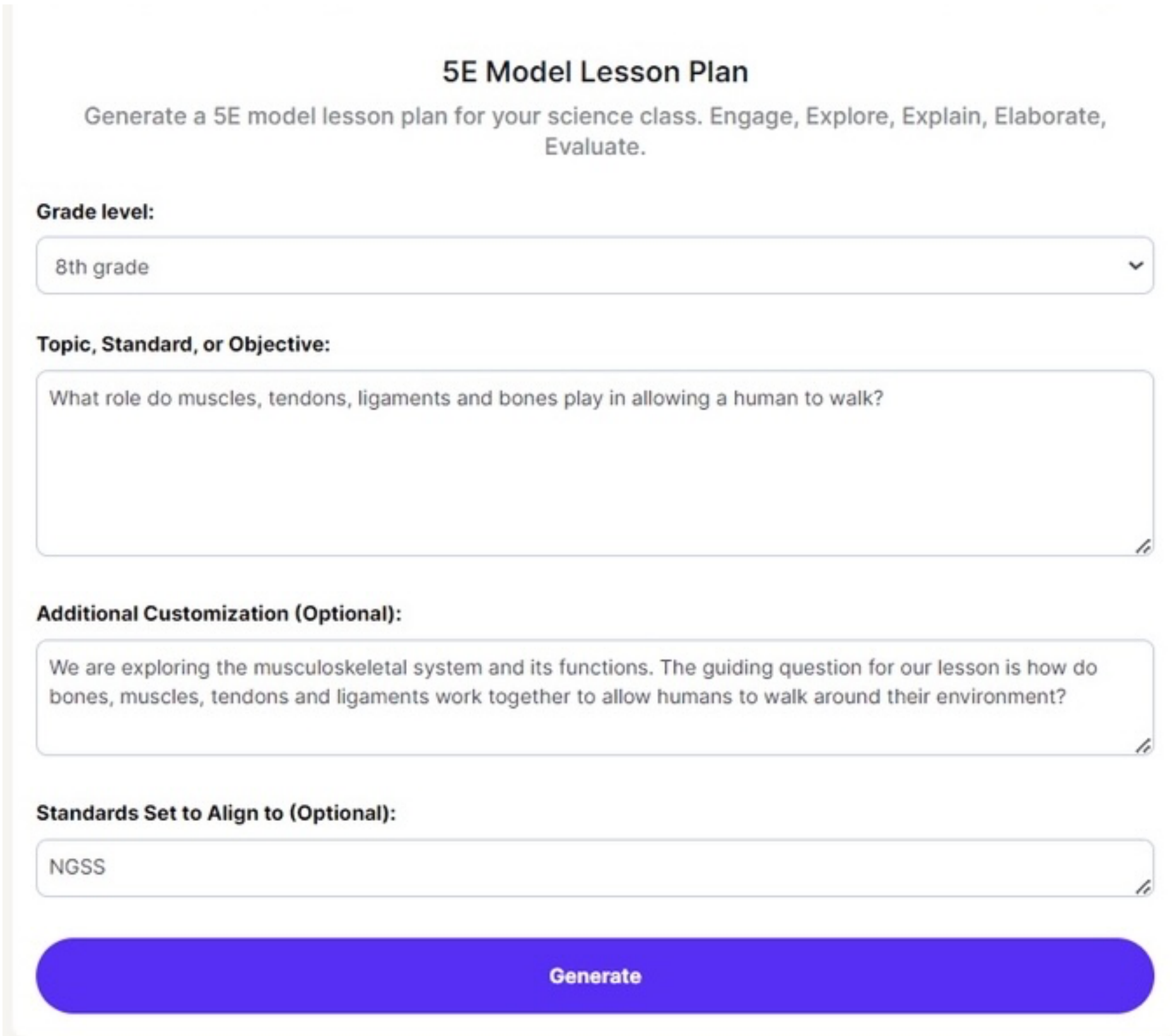

**Fig. 1.** MagicSchool AI Lesson Plan Generator Interface.

*Habit formation.* Sustained use of MagicSchool creates conditions likely to reinforce patterns of quick generation without deep engagement. The platform's value proposition centers on time savings; teachers who invest substantial effort in customization and critical evaluation forfeit the efficiency benefits that motivated adoption. For educators facing documented time pressures, the path of least resistance—generating and accepting "good enough" plans—may become default practice. Over time, this pattern risks diminishing the pedagogical reasoning skills that develop through the effortful work of lesson design. Teachers may increasingly defer to AI-generated structures rather than constructing their own pedagogical logic, potentially eroding the professional judgment that distinguishes expert teaching.

### 4.2 Case 2: Brisk

Brisk is an AI-powered tool that generates feedback on student writing, integrated as a browser extension that works within Google Docs. The platform has achieved substantial adoption, with marketing materials celebrating that teachers can "grade 147 essays in one afternoon" and "reclaim your evenings and weekends." This framing positions assessment as a burden to be minimized rather than a site of professional practice and student connection.

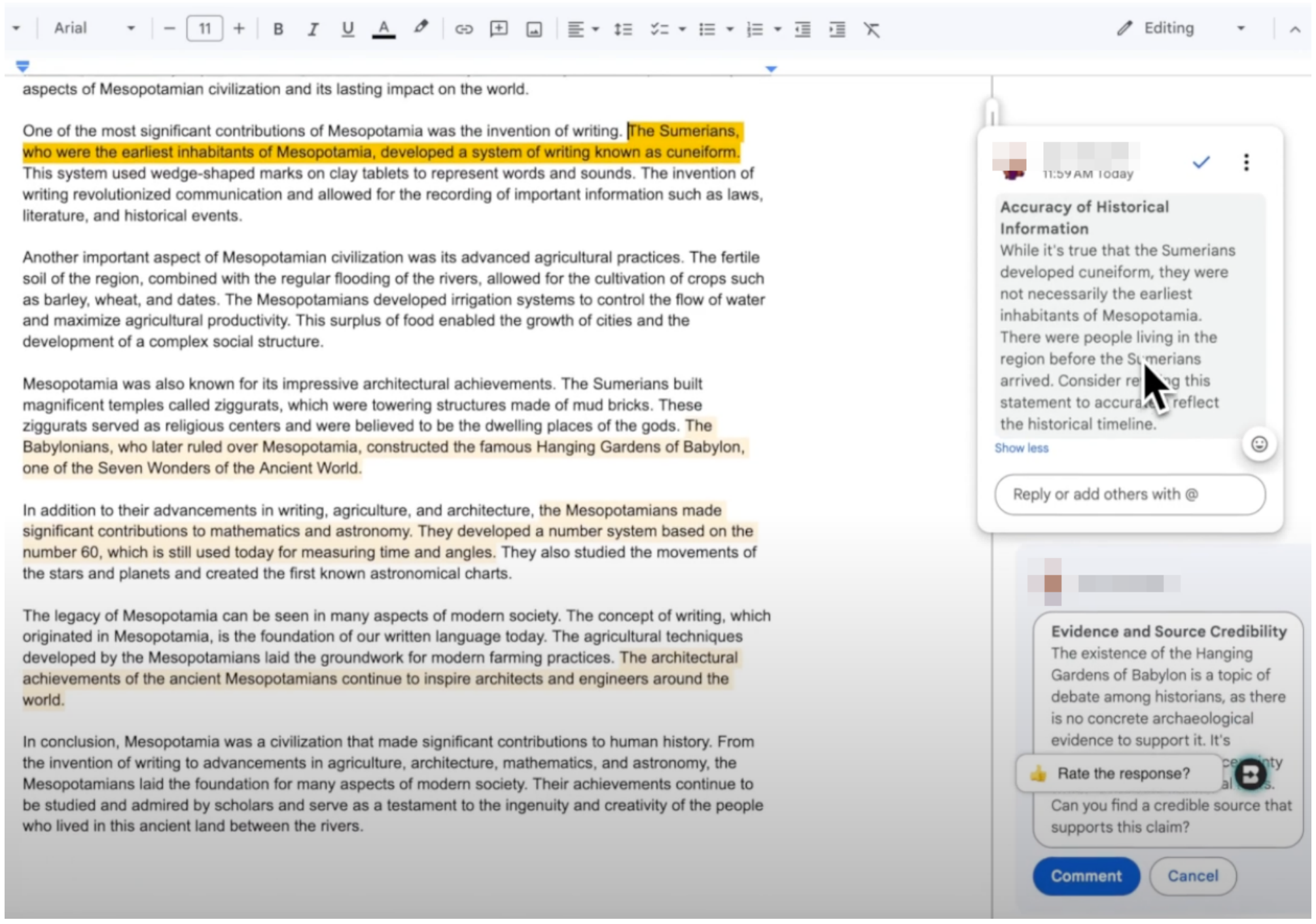

**Fig. 2**. Brisk Automated Feedback Interface.

*Skilled epistemic actions.* Providing feedback on student writing is epistemically rich work. Teachers read student texts, diagnose strengths and difficulties, formulate responses calibrated to individual students' developmental trajectories, and make judgments about which aspects of writing to foreground (Hattie & Timperley, 2007). This process builds teachers' understanding of how students think and learn. Brisk restructures this epistemic labor fundamentally. The AI reads student writing, determines what merits comment, and generates feedback, performing interpretive operations that teachers would otherwise undertake. Teachers can review and modify AI-generated comments, but the platform makes a critical epistemic decision on what aspects of student writing warrant attention. This pre-selection constrains teacher judgment even before the customization stage. The workflow enables processing at scale but transforms teachers from primary readers of student work into reviewers of AI interpretations. Skilled epistemic action remains possible but requires deliberate resistance to the platform's efficiency-optimized design.

*Epistemic sensitivity.* Brisk presents significant challenges across multiple dimensions. First, the AI's determination of what deserves feedback remains opaque, so teachers cannot readily evaluate whether the system's attention aligns with sound pedagogical priorities or their knowledge of particular students. Second, the platform offers no mechanisms for assessing whether generated feedback meets standards for accuracy, constructiveness, or developmental appropriateness. Third, and most significantly, Brisk cannot access the contextual knowledge that informs responsive feedback: students' learning histories, emotional states, prior struggles, and growth trajectories. The system generates comments based solely on textual features, producing feedback that may be generically reasonable but pedagogically decontextualized. Teachers using Brisk must either supply this contextual sensitivity themselves (adding substantial labor that undermines the efficiency rationale) or accept feedback that lacks the relational attunement central to formative assessment.

*Habit formation.* The platform's design creates conditions likely to foster patterns of rapid processing with minimal engagement. Teachers who carefully read each essay, evaluate AI-generated comments against their knowledge of individual students, and substantially revise feedback sacrifice the time savings that constitute Brisk's value proposition. The workflow thus structurally encourages accepting AI interpretations with minimal scrutiny. Sustained use may produce "deskilling" (Selwyn, 2013), as assessment expertise emerges through practice: reading student work closely, identifying patterns, formulating responses, observing effects. When teachers repeatedly delegate the interpretive work of student work to AI systems, the experiential basis for expertise development narrows.

### 4.3 Cross-Case Synthesis

Both cases instantiate epistemic substitution as characterized in the theoretical framework. Rather than augmenting teachers' capacity to perform epistemic operations more effectively, these systems perform operations whose exercise is constitutive of teaching expertise. MagicSchool generates the pedagogical reasoning embedded in lesson structure; Brisk generates the interpretive judgments embedded in feedback. In each case, teachers shift from primary epistemic agents to reviewers of AI outputs, and in doing so, may be prevented from developing the professional expertise that emerges through direct engagement with pedagogical materials and student work.

This substitution operates through consistent mechanisms across both platforms:

- *Efficiency as primary value.* Both systems market time savings as their core benefit, framing the epistemic work of teaching as burdens to be reduced rather than as professional practices with intrinsic value for teacher development and student relationships.
- *Opacity of epistemic operations.* Neither platform makes visible the basis for its AI-generated outputs. Teachers cannot inspect how MagicSchool's pedagogical reasoning behind the generated lesson or how Brisk selects what to comment on. This opacity impedes the critical evaluation that epistemic sensitivity requires.
- *Binary affordance structures.* While both platforms technically permit customization, their interfaces reward acceptance over refinement. The skilled epistemic actions that would maintain teacher agency require working against the platforms' primary design logic.
- *Decontextualized outputs.* Both systems generate outputs based on generic patterns rather than situated knowledge of particular students, classrooms, and communities. The contextual understanding that characterizes teaching expertise—knowing this student's struggles, this classroom's dynamics, this community's values—cannot inform AI generation, yet this knowledge is precisely what distinguishes pedagogically responsive practice from generic knowledge transmission.

These mechanisms reveal how AI systems functioning as epistemic infrastructure can systematically reconfigure professional practice. The individual teacher's choice to customize or accept, scrutinize or defer, is shaped by design decisions that make some patterns of engagement easier than others. When efficiency-oriented design encounters time-pressured practice, the conditions favor epistemic substitution regardless of teachers' intentions. The infrastructure does not determine outcomes, but it tilts possibilities in directions that may erode the epistemic agency through which teaching expertise develops.

**Table 1.** Cross-Case Analysis of Epistemic Conditions in Two AI Systems.

| Condition | MagicSchool AI | Brisk |
|---|---|---|
| Skilled Epistemic Actions | Permits but does not cultivate; interface rewards efficiency over pedagogical refinement | AI pre-selects what merits attention; teachers review rather than generate interpretations |
| Epistemic Sensitivity | Opaque generation; no prompts for evaluation; design rewards acceptance over scrutiny | Opaque selection; no quality mechanisms; decontextualized from student knowledge |
| Habit Formation | Conditions favor quick generation, minimal engagement; risks diminishing pedagogical reasoning | Conditions favor rapid processing; risks algorithmic deskilling of assessment expertise |

## 5. Discussion

The analysis reveals that both AI systems examined structure epistemic possibilities in ways that favor epistemic substitution over augmentation. Teachers using these platforms may shift from performing epistemic operations to approving AI-generated outputs, a reconfiguration with implications extending beyond the specific cases examined.

### 5.1 Contributions to Philosophy of Technology

This analysis advances philosophical understanding of AI systems in three ways. First, it extends postphenomenological frameworks to AI epistemic infrastructure, showing how Ihde's relational categories and Verbeek's concept of composite intentionality illuminate AI's distinctive character as technology that performs cognitive work previously internal to human epistemic processes. The case analyses demonstrate that AI systems do not simply mediate human-world relations in the manner of previous technologies; they reconfigure the epistemic practices through which professional knowledge develops. This extends technological mediation theory to account for systems that substitute for rather than extend human cognitive operations.

Second, the analysis advances epistemic substitution as an analytical concept for understanding AI's impact on human agency. Where augmentation frameworks assume AI enhances human capability, epistemic substitution identifies a different pattern where AI takes over operations that, when performed by humans, are constitutive of expertise development. The concept differs from cognitive offloading generally, as delegating calculation to a calculator raises no such concerns. What makes epistemic substitution problematic is that the specific operations being delegated are those through which professional knowledge forms. This concept provides vocabulary for distinguishing AI systems that genuinely support human epistemic agency from those that supplant it while appearing supportive. Both MagicSchool and Brisk market themselves as supporting teachers, but structural analysis reveals substitution dynamics obscured by efficiency-focused framings but can be detrimental to the profession.

Third, the analysis connects distributed cognition frameworks to virtue epistemology in a way that illuminates AI's implications for professional formation. Theories of extended cognition demonstrate that cognitive processes distribute across human-technology assemblages, but they do not automatically evaluate which distributions support human flourishing. By examining whether AI systems cultivate or erode epistemic virtues (the dispositions of careful evaluation, critical judgment, and epistemic responsibility), this analysis provides normative criteria for assessing human-AI cognitive distributions. The finding that efficiency-optimized AI design may prevent virtue development through practice represents a contribution to understanding how technological mediation shapes epistemic character.

### 5.2 Beyond Education: Structural Epistemic Injustice

The patterns identified in educational AI have broader implications. Herzog and Branford (2025) identify strikingly similar dynamics in medical AI, where decision support systems risk excluding patients and medical personnel from epistemic processes vital to good practice. Physicians, like teachers, develop situated knowledge through direct engagement. They understand a patient's concerns, recognize patterns across cases, and form clinical judgment through sustained practice. AI systems optimized for efficiency threaten to circumvent these processes, positioning professionals as ratifiers of algorithmic outputs rather than primary epistemic agents.

Both cases reveal how AI epistemic infrastructures can create "structural epistemic injustice" that can be understood as systematic exclusion of persons from epistemic processes in which they should participate (Herzog & Branford, 2025). When AI systems perform the interpretive work through which professionals develop situated knowledge, they do not merely change *how* work is done; they reconfigure *who* does epistemically significant work. Teachers become approvers of AI-generated lesson plans rather than authors of pedagogical reasoning; physicians become validators of algorithmic recommendations rather than diagnosticians. This restructuring serves institutional demands for efficiency and scalability but may undermine the epistemic inclusion of professionals and those they serve.

The parallels suggest that epistemic substitution represents a fundamental challenge across professional domains where situated, relational knowledge matters. Law, social work, journalism, design, and other fields where practitioners develop expertise through direct engagement with cases, clients, and contexts may face similar dynamics as AI integration accelerates. The analytical framework developed here, which examines affordances for skilled action, conditions for epistemic sensitivity, and implications for habit formation, illuminates a dilemma between efficiency and epistemic substitution, offering resources for examining AI epistemic infrastructure across these domains. It sensitizes the issue and seeks situated means to address the dilemma.

### 5.3 Design Principles for Epistemic Infrastructure

The analysis suggests normative principles for AI systems intended to support professional epistemic practices.

1. *Preserve conditions for expertise development.* AI systems should be designed to support rather than substitute for the cognitive operations through which professional expertise develops. This requires attending not only to task completion but to the epistemic labor involved in completing tasks. Systems that generate outputs users would otherwise produce may save time while preventing the learning that occurs through production. Design should ask: Does this system help practitioners do epistemic work more effectively, or does it do the work for them?

2. *Support epistemic sensitivity through transparency.* AI systems should make their operations sufficiently transparent that users can exercise critical judgment about outputs. This includes indicating uncertainty, surfacing the basis for selections or recommendations, and providing affordances for evaluation rather than mere acceptance. Design should cultivate rather than suppress users' disposition to scrutinize AI outputs.
3. *Structure interactions to reinforce epistemic virtues.* Interface design shapes habit formation. Systems that optimize for speed and seamless acceptance may foster patterns of uncritical deference; systems that build in moments for reflection, comparison, and deliberation may reinforce epistemic virtues. Design should ask: What dispositions does sustained use of this system cultivate?
4. *Maintain space for contextual knowledge.* Professional practice involves situated knowledge that AI systems cannot access, such as knowledge of a student, a patient, or a local community. AI systems should be designed to complement rather than override such contextual understanding, positioning AI-generated outputs as inputs to professional judgment rather than substitutes for it.

These principles do not reject AI integration but specify conditions under which integration supports rather than undermines human epistemic agency. They are actionable for technology developers who integrate foundation AI models in various contexts. They shift evaluation criteria from accuracy, efficiency, and bias, which are often attributed to AI models, to criteria developers could readily impact such as epistemic sensitivity and professional development.

### 5.4 Limitations and Future Directions

This analysis focuses on teacher-facing AI systems; parallel examination of student-facing systems would extend understanding of how AI epistemic infrastructure shapes learners' epistemic agency and formation. Also, the technical investigation method reveals structural possibilities but not actual use patterns; empirical research examining how teachers engage with these systems in practice would complement and potentially complicate the analysis presented here. Additionally, the analysis addresses mainstream commercial systems; alternative designs that prioritize epistemic agency over efficiency warrant examination to understand whether the patterns identified are intrinsic to AI or artifacts of particular design choices.

The analysis also raises questions beyond its scope. How should we think about the epistemic agency of (agentic) AI systems themselves, if at all? What institutional conditions support designs that prioritize epistemic formation over efficiency? How do power relations—between teachers and administrators, between schools and technology vendors—shape which AI designs proliferate? These questions require further philosophical and empirical investigation.

## 6. Conclusion

AI systems are becoming embedded in epistemic infrastructures across social institutions, reconfiguring the conditions under which humans exercise epistemic agency. This paper developed a framework for analyzing AI systems as epistemic infrastructure and applied it to systems mediating professional teaching practice. The analysis reveals patterns of epistemic substitution, where AI systems perform cognitive operations that are constitutive of expertise development, potentially preventing the development of professional judgment that relies on such practice.

The findings contribute to philosophy of technology by extending technological mediation frameworks to AI systems that substitute for rather than extend human cognition. They contribute to epistemology by connecting distributed cognition to virtue epistemology, showing how infrastructure shapes conditions for epistemic formation. And they contribute to ongoing debates about AI and human agency by specifying mechanisms through which efficiency-optimized design may systematically undermine the epistemic agency it purports to support.

As AI integration accelerates across professional domains, understanding how these systems function as epistemic infrastructure becomes increasingly urgent. The question is not whether to integrate AI but how to design integration that sustains human epistemic agency, the capacity to engage actively in knowledge formation that constitutes professional practice and, ultimately, human flourishing.

## Statements and Declarations

**Acknowledgements** Not applicable.

**Author Contributions** The author confirms sole responsibility for the following: study conception and design, data collection, analysis and interpretation of results, and manuscript preparation.

**Declaration of Generative AI in Writing** During the preparation of this work, the author used large language models (primarily Claude 3.5) to refine the manuscript's clarity, ensure terminological consistency, and format data tables. The author reviewed and thoroughly revised all AI-assisted edits and takes full responsibility for the content of the publication.

**Funding** The authors declare that no funds, grants, or other support were received during the preparation of this manuscript.

**Competing Interests** The authors have no relevant financial or non-financial interests to disclose.

**Ethics Approval** Not applicable.

**Consent to Participate** Not applicable.

**Consent to Publish** Not applicable.